# Optical fiber Sagnac interferometer for sensing scalar directional refraction: application to magnetochiral birefringence.

G. Loas,[a)] M. Alouini, and M. Vallet

*Institut de Physique de Rennes, UMR 6251, CNRS, Université de Rennes 1*

*Campus de Beaulieu, Rennes 35042, France.*

We present a set-up dedicated to the measurement of the small scalar directional anisotropies associated to the magnetochiral interaction. The apparatus, based on a polarization-independent fiber Sagnac interferometer, is optimized to be insensitive to circular anisotropies and to residual absorption. It can thus characterize samples of biological interests, for which the two enantiomers are not available and/or which present poor transmission. The signal-to-noise ratio is shown to be limited only by the source intensity noise, leading to a detection limit of $\Delta\phi = 500$ nrad.Hz$^{-1/2}$. It yields a limit on the magnetochiral index $n_{MC} < 4 \; 10^{-13}$ T$^{-1}$ at 1.55 µm for the organic molecules tested.

## I. INTRODUCTION

In recent years, several interferometric setups have been devoted to the detection of small directional anisotropies, such as, e.g., nonreciprocal circular anisotropies[1], magnetoelectrical directional nonreciprocity[2,3], or magnetochiral anistropy[4,5,6]. These studies aim at a better understanding of the symmetry principles governing the interaction of light and matter and target varied applications. Here, we focus on the magnetochiral interaction, which is important in biochemistry as it has been shown to provide asymmetric photochemical reactions[7] and because of its potential role in the origin of homochirality of life[8]. This fundamental effect consists of a change in the optical index of a chiral media subject to a static magnetic field parallel to the direction of propagation of light[9]. It can be regarded as a cross effect between magnetic optical activity (MOA) and natural optical activity (NOA). As for MOA, its sign depends of the orientation of the magnetic field and, as for NOA, it has opposite sign for the two enantiomers. Moreover, it depends on the direction of propagation of light. But, contrary to MOA and NOA which are circular differential, the magnetochiral

---

[a)] Corresponding author: goulc-hen.loas@univ-rennes1.fr.

interaction is scalar, that is, it does not depend on the polarization of light. The magnitude of this cross effect is weak. For a magnetic field of 5 T and a diamagnetic compound, the amplitude is predicted to be at least $10^{-3}$ smaller than MOA[10]. The first observation of magnetochiral interaction was made in absorption, i.e., magnetochiral dichroism (MCD). Using luminescence spectroscopy techniques, Rikken and Raupach proved that the photo-emission of chiral media depends on the relative orientation of the light with respect to an external magnetic field[4]. This was recently followed by MCD detection in chiral ferromagnets[11] and in organic compounds[12]. For the detection of the refractive part of the magnetochiral interaction, named magnetochiral birefringence (MCB), to the best of our knowledge only two setups have been designed up to now. At Zürich, a folded single-pass Sagnac interferometer was used to detect the variation of directional phase shift associated to MCB[5]. In Rennes, we developed an active interferometer, based on a ring laser, in order to detect the frequency shifts due to intracavity samples[6]. However, for both setups, the response is polarization-dependent. They thus require the availability of the two enantiomers of the compound under study, in order to compensate to zero the circular anisotropies inherent to chiral samples. This drawback prevents from using most of all samples of biological interests, where the availability of the two enantiomers is scare. Moreover, such samples are usually diffusive or absorbing which forbid their insertion into an active interferometer as in Ref. 6. Besides, the experimental results reported in Ref. 5 and 6 do not agree with models based on molecular ab-initio calculations[13]. There is thus a need for additional measurements and consequently for the development of a passive and polarization insensitive interferometer.

Fiber Sagnac interferometer are able to detect very small directional phase-shifts, induced by, e.g., external magnetic fields[14] or mechanical rotation of the interferometer. The latter phenomenon is the so-called Sagnac phase shift[15,] and is the basis of the interferometric fiber-optic gyroscope (IFOG)[16], which has evolved over the three last decades to industrial devices achieving navigation grade performance. Furthermore, in order to minimize the spurious contributions of the fiber anisotropies to the useful signal, a polarization independent technology based on the insertion of depolarizers in the fiber loop has been successfully demonstrated[17,18]. One can then wonder if the detection of small high-order scalar effects such as MCB could benefit from a similar technique by circumventing the contributions of first-order circular anisotropies, here MOA and NOA.

In this paper, we report on the design of a modified fiber-optic depolarized Sagnac interferometer developed in view of investigating MCB at 1.55 µm. In Section II, we detail the setup. Section III is dedicated to the description of the noise sources and to the characterization of possible systematic effects, leading to an estimation of the instrument detection limit. Section IV reports the calibration by means of Fizeau effect. Measurements performed on several organic molecules are presented, leading to a new limit on the magnetochiral birefringence level. Finally, Section V is devoted to the conclusion.

## II. EXPERIMENTAL SETUP

The experimental setup is schematized in Fig. 1. It consists in a table-top fiber-optic Sagnac interferometer operating at 1.5 µm. Part of the loop comprises polarization-maintaining fibers (PM) (grey color in Fig.1), followed by Lyot depolarizers and two sections of standard single-mode fiber (SMF) connected by two collimators. These collimators sandwich the sample under test which presents a non-reciprocal phase shift $\Delta\phi$. The perimeter of the loop is equal to P = 38 m, essentially determined by the length of the depolarizers. The so-called Y-coupler configuration of the interferometer is composed of a $LiNbO_3$ integrated optic circuit (IOC) provided by Photline Inc. This multifunction circuit, originally designed for IFOG, integrates a 50/50 coupler, a push-pull phase modulator and a 60 dB extinction ratio polarizer on a Lithium Niobate waveguide. The $LiNbO_3$ guides are fabricated using proton exchange technique.

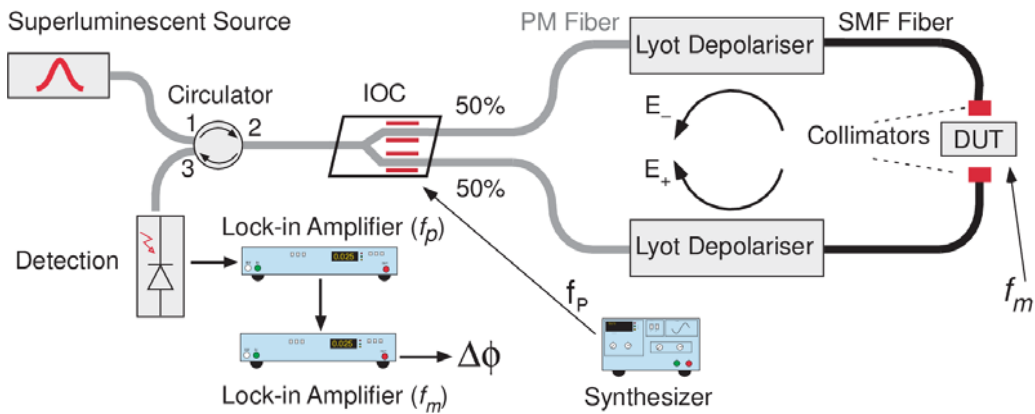

Fig. 1: Scheme of the modified depolarized Sagnac interferometer. See text for details.

The optical source is a 2 mW superluminescent diode at λ = 1.55µm (Superlum Inc.). As discussed abundantly in the literature[19,20], broadband emission is mandatory in order to avoid residual interferences that might be caused by coherent back-reflections and backscattering in the interferometer. It also permits to get rid of spurious non reciprocities due to Kerr nonlinearities, i.e., a non-reciprocal change of optical index due to slight power imbalance between the two counterpropagating waves. Moreover, as will be detailed later, a broadband source is required to efficiently depolarize the two counterpropagating optical beams travelling through the sample under test. As shown on Fig. 2(a), the spectral width of our source is measured to be equal to $\Delta\lambda = 60$ nm, leading to a coherence time $\tau_c = 130$ fs and a coherence length of 26 µm in the fiber. A superluminescent diode is preferred because its emission spectrum presents a Gaussian shape which optimizes the flatness of the degree of coherence $\Gamma(\tau)$ for $\tau > \tau_c$, as reported in Fig. 2(b). An optical circulator permits to direct the beam to the interferometer and extract the output beam for detection.

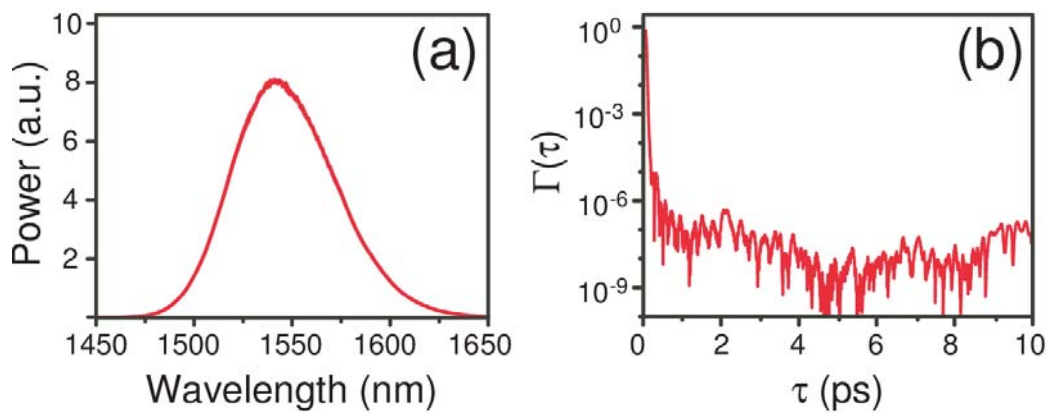

Fig. 2: (a) Optical spectrum of the broadband source and (b) associated coherence function.

The two in-loop home-made Lyot depolarizers[21] are the key elements for our Sagnac interferometer. They consist of two sections of Panda birefringent fiber spliced at 45° with respect to each other, with associated length L and 2L and 4L and 8L respectively (see Fig. 3). In our setup, L is equal to 1.5 m.

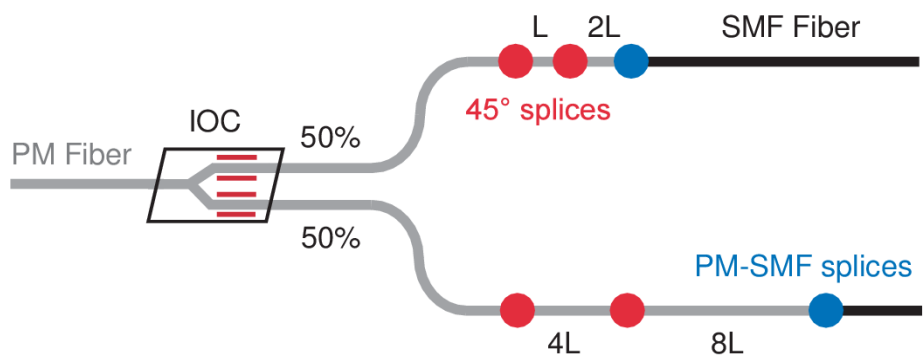

Fig. 3: Schematic of the 2 pieces Lyot depolarizer inserted in the Sagnac interferometer.

As discussed by Szafraniec[18], the depolarized interferometer drastically reduces the contribution of in-loop spurious non-reciprocal linear and/or circular birefringences $\Delta\phi_L$ and $\Delta\phi_C$. Indeed, in the case of a linear birefringence, the effective birefringence $\Delta\phi_L^{eff}$, i.e., detected by the interferometer, is given by

$$\Delta\phi_L^{eff} = (\Delta\gamma)^n \Delta\phi_L, \tag{1}$$

with $\Delta\gamma$ the misalignment angle between fiber splices and n the number of sections in the depolarizers. Assuming that the mean misalignment is lower than 1°, it yields $\Delta\phi_L^{eff} < 3.10^{-4} \Delta\phi_L$. In the case of a circular birefringence, the effective birefringence $\Delta\phi_C^{eff}$ measured by our Sagnac interferometer and related to the non-reciprocal circular birefringence reads

$$\Delta\phi_C^{eff} = \left(\Gamma(\delta\tau_1) + \Gamma(\delta\tau_2)\right) \Delta\phi_C / 2. \tag{2}$$

$\delta\tau_1$ and $\delta\tau_2$ are the group delay differences associated to the two depolarizers, denoting difference of time propagation between the proper axes of the birefringent fiber. The beat length of the panda fiber we used is of the order of two millimeters[22], leading to $\delta\tau_1$ and $\delta\tau_2$ equal to about 3 ps and 12 ps, respectively. From Fig. 2(b), the associated degrees of coherence $\Gamma(\tau_1)$ and $\Gamma(\tau_2)$ are then shown to be lower than $10^{-6}$. This means that $\Delta\phi_C^{eff} < 10^{-6} \Delta\phi_C$.

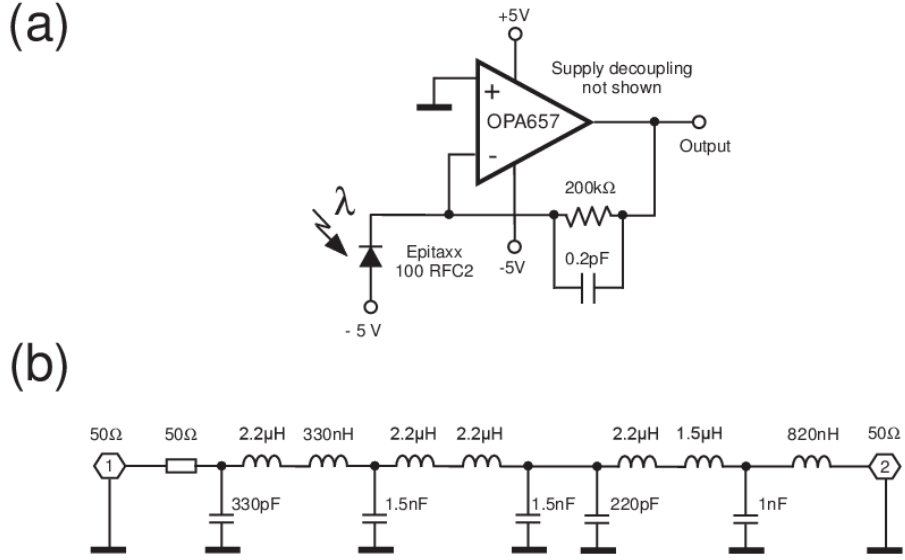

Fig. 4: (a) Transimpedance amplifier. (b) 8$^{th}$-order Butterworth filter.

The mean power impinging on the photodiode is of the order of 10 µW. The photodiode is followed by a low-noise transimpedance amplifier whose electrical diagram is reported in Fig. 4(a). This assembly is positioned in a Faraday box in order to isolate it from parasitic signals. In order to detect the in-loop directional phase shift $\Delta\phi$, the interferometer must be biased at an operating point with a non-zero response slope. As in usual FOG, an asymmetric phase modulation is applied at a frequency matching the loop proper frequency defined by $f_p = 1/(2\tau_g)$, with $\tau_g$ the propagation time in the Sagnac ring. In our setup, $f_p$ is equal to 2.7 MHz. This phase modulation of amplitude $\phi_p$ offers two major advantages. First, it permits a heterodyne phase measurement around the offset frequency $f_p$, thus rejecting low frequency noises. Second, it is applied at the entrance of one arm of the interferometer only, leading to a phase bias at the detection side between the two contra-propagating waves[19]. Actually, the output signal from the transimpedance amplifier can be written as

$$V(t) = V_0 \left[ 1 + \cos\left(\Delta\phi + \phi_p \cos(2\pi f_p t)\right) \right], \qquad (3)$$

where $V_0$ depends of the transimpedance gain and $\phi_p$ is the phase modulation amplitude. By using a standard Bessel decomposition and assuming $\Delta\phi \ll 1$, the amplitudes of the component of the signal at $f_p$ and at $2f_p$ are respectively equal to

$$A(f_P) = 2V_0 J_1(\phi_p) \sin \Delta\phi \cong 2V_0 J_1(\phi_p) \Delta\phi, \qquad (4.a)$$

$$A(2f_P) = 2V_0 J_2(\phi_p)\cos\Delta\phi \cong 2V_0 J_2(\phi_p). \tag{4.b}$$

The non-reciprocal phase shift of interest can be easily deduced from the ratio of these two components as

$$\Delta\phi = \left(J_2(\phi_p)/J_1(\phi_p)\right)\left(A(f_P)/A(2f_P)\right). \tag{5}$$

Experimentally, $A(2f_P)$ is first measured by a lock-in amplifier at $2f_P$ whose integration time is set to 30 ms. In order to extract $A(f_P)$, an analog 8$^{th}$ order Butterworth low-pass filter[23] (detailed in Fig. 4(b) is used before lock-in amplifier detection, now at $f_P$. The −3dB cut-off frequency of the filter is equal to 3.5 MHz, leading to a rejection level of 30 dB at $2f_P$. We have experimentally noticed that such a high rejection level was mandatory in order to avoid saturation of the input stage of the lock-in amplifier by the $2f_P$ component.

As discussed in the next section, the amplitude of the directional phase-shift of interest has to be also modulated in order to increase the detection sensitivity. However, this modulation must be done at a very low frequency $f_m$, that is, well below the time response of the lock-in amplifier which is ruled by its integration time (30 ms in our case). – When the non-reciprocal effect of interest is for example the magnetochiral birefringence, this modulation can be performed through the amplitude of the magnetic field. Now assuming that $\Delta\phi = \phi_m \sin(2\pi f_m t)$, a second lock-in detection at $f_m$ will provide the amplitude of the components at $f_P \pm f_m$ that are given by $A(f_P \pm f_m) = 2V_0 J_1(\phi_p)\phi_m$ using Eq. (4.a). $\phi_P$ is experimentally adjusted to 1.8 rad in order to maximize this signal. From Eq. (4) and Eq. (5), one gets

$$\phi_m = 0.53\ A(f_P \pm f_m)/A(2f_P). \tag{6}$$

A double modulation-demodulation scheme will thus enable the detection of a modulated non-reciprocal effect. Here, the first lock-in amplifier (high frequency) is a Stanford Research SR844, whereas the second one (low frequency) is a 7220 from EG&G. Let us mention that the selected output filter of the latest is a the fourth-order low pass filter, maximizing the noise rejection to an equivalent bandwidth of 5/(64T) with T the integration time.

Finally, we point out the fact that large phase dynamic range is here not required, since the amplitude of the non-reciprocal phase shift is supposed to be fairly constant during the measurement time. It is thus not necessary to apply a closed-loop signal processing, as in standard FOG for which large dynamic range is obtained by using feedback electronics driving a control element in the interferometer[19].

## III. DETECTION LIMIT

In this section, we evaluate the noise floor of the detection and discuss the amplitudes of the spurious non reciprocal offsets that could add to the signal.

The noise floor $\Delta\phi_N$ associated to $\Delta\phi$ is equal to the ratio of the current noise $\langle i_N^2 \rangle^{1/2}$ with respect to the photocurrent $\langle i_{ph} \rangle$, both measured at the output of the Butterworth filter. This ratio is reported in Fig. 5(a). At the detection frequency $f_P$, the noise floor value of the interferometer is about 0.5 µrad/Hz$^{1/2}$. This level is comparable to the noise floor obtained in state-of-the-art FOG[24]. At $f_p = 2.7$ MHz and for an integration time of 1000 s, the detection limit of our setup is potentially of 5 nrad. This has to be compared to the photon and electron noise floor. This later is obtained from the root sum squaring of shot noise, relative-intensity noise (RIN), and Johnson noise of the load R, that is[20]

$$\Delta\phi_N = \langle i_N^2 \rangle^{1/2} / \langle i_{ph} \rangle = \left(2e\langle i_{ph}\rangle + 1.38 \langle i_{ph}\rangle^2 / \Delta\nu + 4k_B T / R\right)^{1/2} \Delta f^{1/2} / \langle i_{ph} \rangle, \qquad (7)$$

with $\Delta f$ the detection bandwidth, and where the factor 1.38 stands for phase-to-intensity conversion in depolarized Sagnac interferometers. As can be seen on Fig. 5(b), its value corresponds to the experimental level. It shows that, here, the detection limit is ruled by the RIN of the broadband source.

Static non reciprocal phase shifts lead to a DC output voltage at the output of the first lock-in amplifier, which adds to the signal of interest at $f_m$. Their contributions may spoil or even overcome the useful signal, depending on the rejection level of the second lock-in amplifier, and thus have to be minimized. First, the spurious directional linear and circular anisotropies should be reduced by four and six orders of magnitude respectively, as discussed in Section II. In particular, we have experimentally verified the sensitivity of our depolarized Sagnac interferometer to the Faraday effect. A 0.13 T magnetic field was applied on a 1 cm section of SMF fiber in the loop. The Verdet constant of silica being of about 0.6 m$^{-1}$T$^{-1}$, it gives a Faraday phase shift of 0.8 mrad. No signal was detected at $f_p$. The effect is thus smaller than the detection noise given by (7). This confirms that our interferometer permits us to get rid of vectorial effects in the limit of a depolarization better than $10^{-5}$. Second, contrary to FOG, no particular effort has been performed on the winding of the fiber to minimize possible asymmetric thermal drift of the fiber. This effect, known as shupe effect[25], is here negligible since the fiber loop is extremely short as compared to FOG. Then, Rayleigh backscattering (and back reflections on the collimators) might also provide non reciprocal noise. Its contribution can be circumvented by symmetrizing the loop[26]. We have therefore placed the two collimators at equal distance from interferometer entrance. Finally, the tiny Sagnac directional phase shift due to earth rotation was minimized by properly orienting the sensing loop plane perpendicular to the terrestrial rotation vector. In summary, from the DC component of the first lock-in output and using (5), the total static phase shift was measured to be lower than 1 µrad. No contribution of this spurious phase shift to the output signal of the second lock-in, i.e., at $f_m$, was detected.

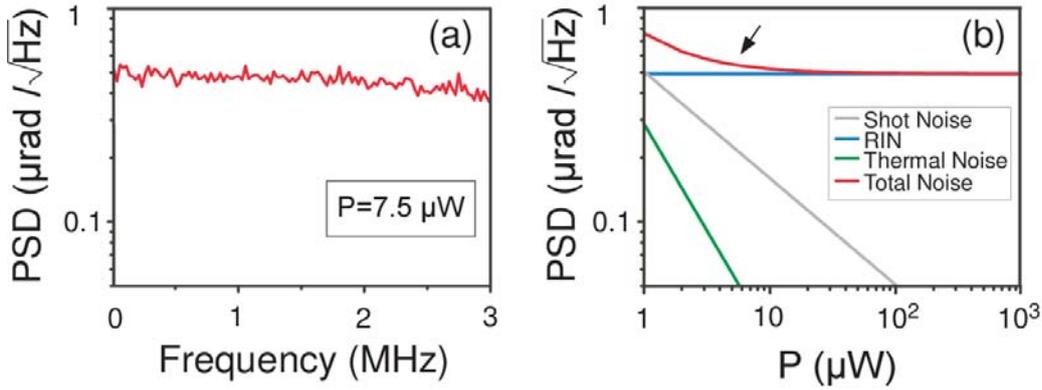

Fig. 5: (a) Experimental power spectral density (PSD) versus frequency. (b) Estimated PSD versus optical power P impinging on the photodiode. The arrow indicates optical power in our experimental conditions (7.5 µW).

Since our interferometer was specifically designed for sensing small modulated non-reciprocal phase shifts, that is scalar directional refraction, a possible small modulated scalar absorption associated to the effect under study may contribute to the signal of interest. In order to give an order of magnitude of this contribution and without loss of generality, we can write the counterpropagating optical fields $E_\pm$ (see Fig. 1) as

$$E_\pm(t) = E_0 \left[1 + (\varepsilon \pm \Delta\varepsilon)\sin(2\pi f_m t)\right] \exp\left[\pm j(\phi_m \sin(2\pi f_m t) + \phi_p \cos(2\pi f_p t))/2\right], \qquad (8)$$

with $E_0$ the amplitude of the fields and where $\varepsilon$ and $\Delta\varepsilon$ correspond respectively to the reciprocal and non-reciprocal part of the absorption modulated at $f_m$ within the component under test. Instead of (3), the output voltage is now equal to

$$V(t) = 2V_0 \left[1 + \varepsilon \sin(2\pi f_m t)\right]^2 \left[1 + \cos(\phi_m \sin(2\pi f_m t) + \phi_p \cos(2\pi f_p t))\right]$$
$$+ 2V_0 \Delta\varepsilon^2 \sin^2(2\pi f_m t) \times \left[1 - \cos(\phi_m \sin(2\pi f_m t) + \phi_p \cos(2\pi f_p t))\right]. \tag{9}$$

At the first order with respect to $\varepsilon$, $\Delta\varepsilon$ and $\phi_m$, expression (9) simplifies to

$$V(t) = 2V_0 \left(1 + 2\varepsilon \sin(2\pi f_m t)\right) - 2V_0 \phi_m \sin(2\pi f_m t) \sin(\phi_p \cos(2\pi f_p t))$$
$$+ 2V_0 \left(1 + 2\varepsilon \sin(2\pi f_m t)\right) \cos(\phi_p \cos(2\pi f_p t)). \tag{10}$$

A straightforward calculation shows that both components $A(f_p \pm f_m)$ and $A(2f_p)$ depend neither on $\varepsilon$ nor on $\Delta\varepsilon$. Relation (6) thus still applies when $\phi_p$ is adjusted to 1.8 rad. Consequently, at first order, the double modulation-demodulation scheme that we propose cancels any spurious contribution related to an absorption term modulated at the same frequency that the refraction term under study. This is confirmed by the measurements that are now detailed in the next section.

## IV. EXPERIMENTAL RESULTS

In order to check the calibration of our interferometer, we have first measured the tiny scalar directional phase shift associated to the well-known Fresnel drag effect. When two beams counterpropagate in a medium of length $L_0$ moving along the propagation axis, they experience a directional phase shift $\Delta\phi_{FD}$ given by the following formula[27]

$$\Delta\phi_{FD} = \left(n - 1 - \lambda \frac{\partial n}{\partial \lambda}\right) \frac{4\pi v L_o}{\lambda c}, \tag{11}$$

where v is the displacement speed of the medium and n its refractive index. Here, we took as moving medium a 51 mm long cylinder of fused silica that was inserted between the two collimators. It was periodically translated back and forth at $f_m = 0.3$ Hz by means of a motorized stage. Figure 6 reports the measured phase shifts obtained when v was varied from 0.25 mm.s$^{-1}$ to 25 mm.s$^{-1}$. Perfect agreement is observed between the experimental data and the theoretical expectations from (11). These results evidence the good sensitivity and linearity of the interferometer response. Let us mention that the moving glass rod induces slight reciprocal losses that are periodically modulated at $f_m$ because of unavoidable small misalignments. The associated amplitude was estimated to be in the range of $\varepsilon \approx 10\%$. Nevertheless, as expected from (10), it does not contribute to the signal, which evidences the insensitivity of the setup to modulated absorption.

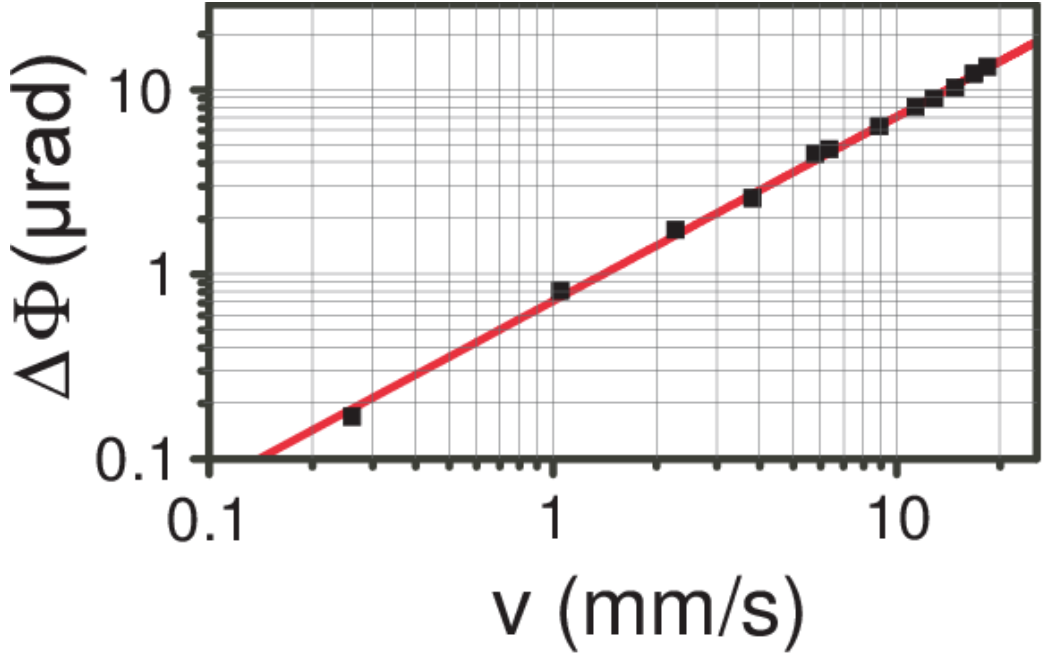

Fig. 6: Calibration of the interferometer with using Fresnel-Drag phase shift.

The experimental arrangement proposed to detect the magnetochiral birefringence consists of a $L_c = 1$ cm-long fused silica cell filled with the sample under study, as shown on Fig. 7. A Nd-Fe-B permanent magnet is mounted on a mechanical stage rotating at $f_m = 1$ Hz. The rotation of this magnet induces along the optical axis an amplitude modulation of the magnetic field H of 0.13 T. When an achiral compound, such as acetone, is poured inside the cell, we detect a residual phase shift that follows the rotation of the magnet. We found that this systematic signal corresponds to a cross effect between the linear strain birefringence $\phi_L$ of the cell windows and the rotation of plane of polarization $\theta_F$ experienced by the optical beam through the sample, i.e., circular birefringence associated to the Faraday effect. Experimentally, the amplitude of the phase shift varies from a few tens to a few hundred of nrad, depending on the spot position on the cell window, that is, on the value of the probed residual birefringence. If the input beam was linearly-polarized, the amplitude of this phase shift would have been of the order of $\phi_L \theta_F$ and dependent of the direction of polarization with respect to the birefringence neutral axes[28]. The typical residual strain birefringence of a 1mm-thick silica window is of the order of 5 mrad[29], while the Faraday rotation angle $\theta_F$ is estimated to be 6 mrad. Consequently, a phase shift of the order of 30 µrad is expected from theory for a linearly polarized beam. The measured systematic effect is well below this value which is consistent with the fact that the beam travelling through the cell windows is depolarized. This is confirmed by Fig. 8, which reports the systematic phase shift when a quarter wave plate, i.e., $\phi_L$ = 1.5 rad, is inserted between the cell and one collimator for several orientations. Although $\phi_L \theta_F$ is now equal to 9 mrad, the measured signal never exceeds 1.2 µrad. The drastic reduction of this systematic effect is made possible because the sensing beams are depolarized. This justifies the need of perfectly depolarized beams and consequently the use of a passive interferometer.

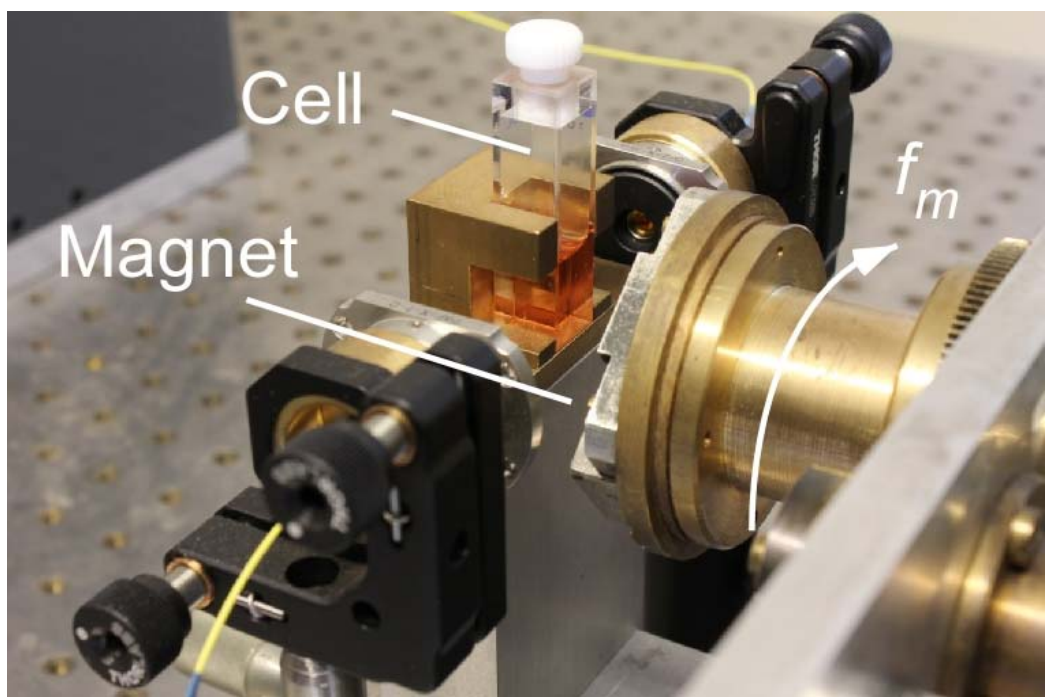

Fig. 7: Setup of the magnetochiral measurement. Scale: the cell is 1cm-long.

Measurements on chiral compound can be now conducted. To compare with previous measurements, we consider samples formerly tested at 488nm with an active interferometer[5] and at 633 nm with passive interferometers[6]. However, a rough $\lambda^{-2}$ dependence for the magnetochiral index is predicted, both from dipole-dipole interaction model[30] and from full quantum-mechanical considerations[31]. The expected effect is thus lower at 1550 nm compared to visible wavelengths and a large rotatory dispersion is required in order to get a detectable magnetochiral effect. Three molecules comply pretty well with the previous requirements: limonene, 3-(trifluoroacetyl)-camphor and carvone. These three molecules are fairly transparent at 1550 nm, both enantiomers are available and the optical activity is rather large. The first measurement was done with racemic limonene. A systematic phase shift of 30 nrad was detected. We have then successively replaced the racemic mixture by R(+)-limonene and S(−)-limonene. Even with large integration times (1000s) on the lock-in amplifier, corresponding to a noise rejection equivalent bandwidth $\Delta f = 0.1$ mHz, we did not detect any magnetochiral directional phase shift within 10 nrad fluctuations.

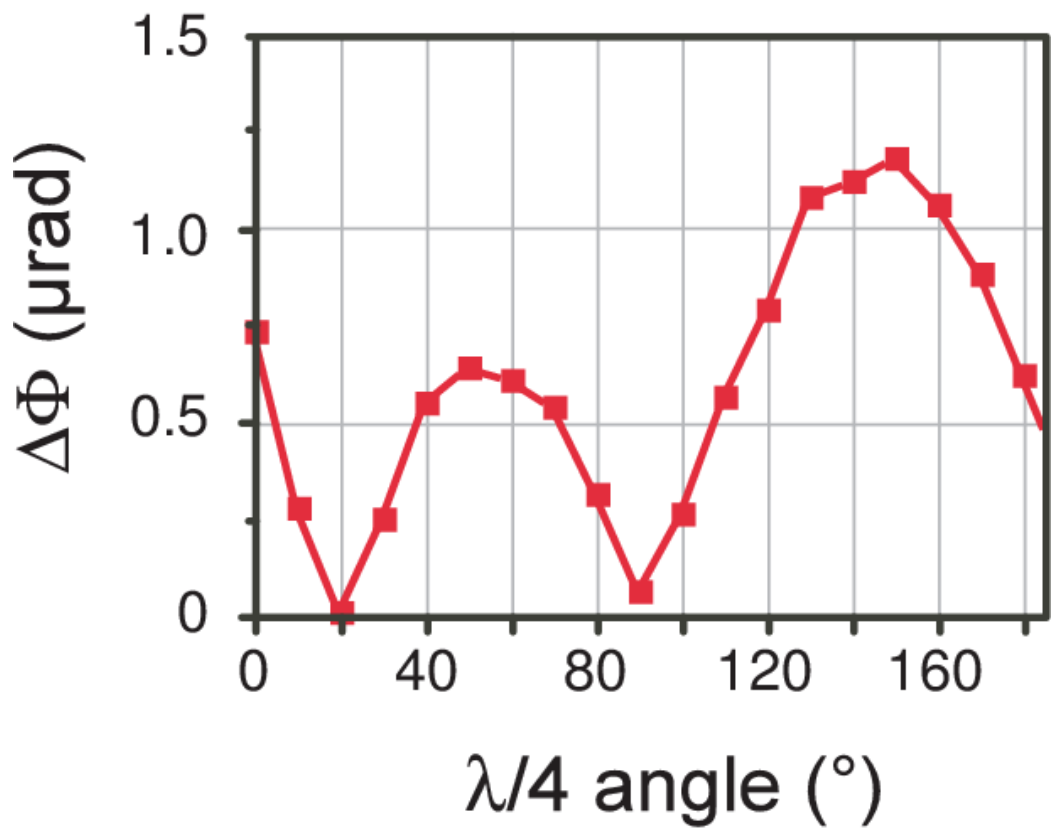

Fig. 8: Systematic phase-shift induced by an in-loop quarter wave plate.

By considering that a signal-to-noise ratio of 3.3 avoids false alarm and false dismissal (probability better than 99.6% for a Rayleigh distribution), we can thus affirm that $\Delta \phi_{MC} < 33$ nrad. Defining the magnetochiral index as $n_{MC} = (\Delta \phi_{MC} \lambda)/(4\pi L_C)$, it yields a limit $n_{MC} < 4\ 10^{-13}$ T$^{-1}$. Similar negative results were obtained on carvone and 3-(trifluoroacetyl)-camphor.

Assuming absorption bands in the far UV, an order of magnitude of the expected phase shift $\Delta \phi_{MC}$ can be obtained from an expression derived from the Bequerel relation extended to magnetochiral interaction[10,28].

$$\Delta\phi_{MC} = \frac{eH}{2\pi mc} \frac{\lambda \lambda_0^2}{\left(\lambda^2 - \lambda_0^2\right)^2} \theta_A, \qquad (12)$$

where e and m are the charge and mass of the electron and where the angle of rotation due to optical activity obeys the following wavelength dependence $\theta(\lambda) = \theta_A / \left(\lambda^2 - \lambda_0^2\right)$. For Limonene, $\theta_A$ and $\lambda_0$ are estimated to be respectively 600°/dm and 210 nm. Eq. (12) then leads to $\Delta\phi_{MC}$ = 500 nrad, that is $n_{MC}$ = 4.7 $10^{-11}$ T$^{-1}$. This is two orders of magnitude above the noise floor of our instrument. . This can also be compared to the values measured at visible wavelength. For limonene, we measured in the past[6] $n_{MC}$ = 3.9(±1.3) $10^{-10}$ T$^{-1}$ at 488 nm, while at 633 nm, Kleindienst and Wagnière measured, for 3-(trifluoroacetyl)-camphor and for carvone, respectively $n_{MC}$ = 3(±0.2) $10^{-8}$ T$^{-1}$ and $n_{MC}$ = 1.3(±0.3) $10^{-9}$ T$^{-1}$ [5]. Again, a $\lambda^{-2}$ dependence for the magnetochiral index then leads to expected estimations at 1.55 µm two orders of magnitude above detection limit of our instrument. These unexpected results might mean that the $\lambda^{-2}$ dependence has to be reconsidered or/and that the magnetochiral index is actually much lower than that expected from Bequerel model.

## V. CONCLUSION

An apparatus combining a depolarized fiber-optic Sagnac interferometer and a double modulation-demodulation scheme has been designed to measure the non-reciprocal phase shifts associated to the magnetochiral index. We have validated a depolarization level better than $10^{-5}$. This permits to decrease the amplitude of systematic phase shifts, mainly due to residual birefringences of the cell windows, below 30 nrad. The good stability of the interferometer allows one to reach a measurement time of 1000 s with a noise-floor value of 0.5 µrad/Hz$^{1/2}$ comparable with the state-of-the-art. This yields an experimental detection level of 33 nrad with a confidence level of 99.6%. Tests on three different organic molecules have shown that the magnetochiral index is lower than 4 $10^{-13}$ T$^{-1}$ at 1.55 µm. As compared to figures previously obtained for the same molecules in the visible[5,6], it implies a significant discrepancy with the values expected from a $\lambda^{-2}$ dependence for the magnetochiral index. To validate our results, the next step is consequently to compute accurately the magnetochiral index at 1.55 µm from ab-initio models[13]

As the detection performance reaches the noise floor defined by the relative intensity noise of the superluminescent source, it seems difficult to drastically lower the limit of this setup below 33 nrad, the acquisition time being already set to its maximum. Increasing this acquisition time further would only slightly diminish the noise floor, at the expense of a larger sensitivity to thermal and mechanical long-term fluctuations. In order to detect magnetochiral index with our setup, a possibility would be to use larger alternated magnetic field[4]. Alternatively, the availability of samples that present larger magnetochiral birefringence is an open question. Indeed, compared to resonant passive ring interferometers[3,32], our single-pass setup is not sensitive to samples which present residual absorption or diffusion. One can then consider using samples made of chiral compounds mixed with ferrofluid[30], or chiral ferromagnets[11]. Organic compounds with a large optical activity in the infrared, due to, e.g., delocalized π-electrons such as helicene, could also be tested[33]. The apparatus could also be easily modified to test in-loop samples under reflexion at non-normal angle. This would permit to test solid samples, such as, e.g., photonics crystals[34] where giant magnetochirality is expected. Perspective for the setup would also include the extension to the detection of other magnetical directional anisotropies in crystals[35]

**Acknowledgments**

The authors thank Elsa Rescan for preliminary works, Cyril Hamel and Ludovic Frein for technical assistance, Thierry Ruchon and Jeanne Crassous for fruitful discussions. This work was partly funded by Université de Rennes 1.